\documentclass[pdflatex,sn-mathphys-num]{sn-jnl}

\theoremstyle{thmstyleone}%
\theoremstyle{thmstyletwo}%

\theoremstyle{thmstylethree}%

\newcommand{\thth}{\theta\theta}
\newcommand{\phph}{\varphi\varphi}
\usepackage{amssymb,amsmath}
\usepackage{bm}
\usepackage{mathrsfs}

\DeclareRobustCommand{\Erase}{\bgroup\markoverwith{\textcolor{red}{\rule[.5ex]{2pt}{0.4pt}}}\ULon}
\newcommand{\linea}{\textit{linea }}
\newcommand{\muCT}{$\mu$CT}
\usepackage{amsmath}

\usepackage{pdfpages}

\begin{document}
\vspace{-3cm}
\title[Crack on shells]{Where Humpty Dumpty Breaks:\\ Geometry-Driven Fracture in Ellipsoidal Shells}


\author[1]{\fnm{Naoki} \sur{Sekiya}}
\author[2]{\fnm{Yuri} \sur{Akiba}}
\author[3]{\fnm{Kai} \sur{Kageyama}}
\author[3]{\fnm{Hokuto} \sur{Nagatakiya}}
\author[3]{\fnm{Ryuichi} \sur{Tarumi}}
\author*[1]{\fnm{Tomohiko} G. \sur{Sano}\email{sano@mech.keio.ac.jp}}

\affil[1]{\orgdiv{Department of Mechanical Engineering}, \orgname{Keio University}, \orgaddress{\city{Yokohama}, \postcode{2230061}, \state{Kanagawa}, \country{Japan}}}

\affil[2]{\orgname{Mount Fuji Research Institute, Yamanashi Prefectural Government}, \orgaddress{\city{Fujiyoshida}, \postcode{4030005}, \state{Yamanashi}, \country{Japan}}}

\affil[3]{\orgname{Graduate School of Engineering Science, Osaka University}, \orgaddress{\street{1-3 Machikaneyama}, \city{Toyonaka}, \postcode{5608531}, \state{Osaka}, \country{Japan}}}


\abstract{
Fracture networks are ubiquitous in nature, spanning scales from millimeter-sized cracks in botanical peels to hundred-kilometer-long lineae on planetary satellites~\cite{Karpowicz1989, Corder2005, Gordon2009structures, Mallet1875, Akiba2021, Europa}. 
The propagation of a crack is a complex, nonlinear phenomenon governed by the interplay of mechanical properties, rheological behavior, and system geometry.
While fracture mechanics has long addressed structural failure, the relationship among fracture, elasticity, and nonlinear geometry has recently revived as a focal point in condensed matter and biophysics. However, a unified framework that systematically explains how surface geometry prescribes the transition between disparate fracture morphologies remains elusive.
Here we show that shell curvature provides a geometric blueprint for fracture, governing the evolution of complex crack networks through induced stress anisotropy. 
By internally pressurizing thin, bilayer spheroidal shells, we demonstrate that a rich diversity of crack morphologies across lateral, longitudinal, and random orientations depends on the curvature ratio between the pole and the equator. We find that these patterns arise from the nonlinear mechanics of the shell, which can be leveraged to effectively control crack growth.
Our results establish a direct link between structural curvature and fractures, providing a predictive framework that integrates nonlinear geometry with the classical Griffith and von Mises criteria~\cite{Anderson2005fracture}. 
Beyond our model system, we find that the disparate fracture patterns observed in ripening muskmelons and in the icy crust of Europa follow the same geometric principles~\cite{Gerchikov2008, Akiba2022, Europa}. We expect that this unified understanding of crack morphogenesis will inform the design principles of novel functional materials that are resilient to fracture~\cite{Nam2012, Kim2021, Thorimbert2024} and provide insights into the mechanical performance of curved biological and geophysical architectures.
}




\maketitle

\section{Introduction}\label{sec1}

\begin{figure}[!h]
    \centering
    \includegraphics[width =\textwidth]{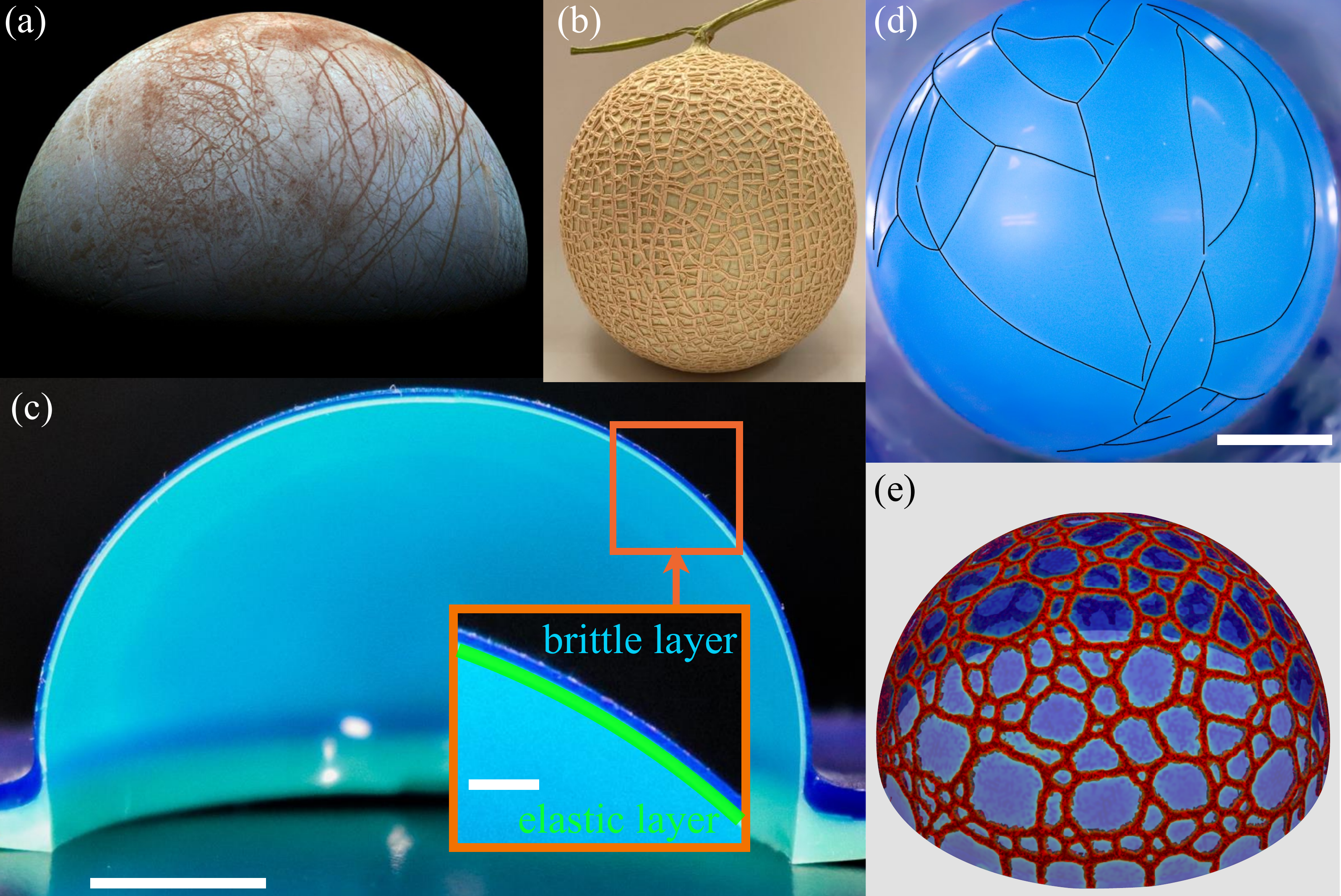}
    \caption{\textbf{Cracks on Europa, melon surface, and model bilayer shells} (a) Surface image of the moon of Jupiter, Europa, made from images taken by NASA's Galileo spacecraft in the late 1990s. (b) Photograph of a cantaloupe, (c) Model experimental system of fracture on shells. Cross-section of bilayer spheroidal shell of the inner-elastic and outer-brittle layers (scale bar: 10 mm). The inset shows a close-up view of the cross-section of the bilayer shell (scale bar: 1 mm). (d) Top view of the post-fractured surface of one of the shells (as Fig.~\ref{fig:2}(c-iii)). (e) Simulation results from the phase-field-combined FEM simulation.}
    \label{fig:1}
\end{figure}

Fracture networks emerge across an immense range of length scales, from the millimeter-sized craquelure of aging paint~\cite{Karpowicz1989, Corder2005}, structures and arches~\cite{Gordon2009structures}, kilometer-sized columnar jointing~\cite{Mallet1875, Akiba2021} and even to the hundred-kilometer-long \textit{lineae} observed on the icy crust of Europa (Jupiter II), one of the satellites of Jupiter (Fig.~\ref{fig:1}(a))~\cite{Europa}. 
Europa is primarily composed of silicate rock and has a water-ice crust and, probably, an iron-nickel core. Huge tidal forces from Jupiter or tectonic forces work on the icy crust of $\sim$100km, leading to numerous characteristic line-shaped patterns such as \textit{linea}~\cite{Europa}.
While the failure of materials and structures has been studied extensively to avoid catastrophic failure, predicting and controlling specific crack trajectories remains a formidable challenge~\cite{Nam2012, Kim2021, Thorimbert2024}. Although classical frameworks like the Griffith and von Mises criteria define the onset of failure~\cite{Anderson2005fracture}, the fundamental difficulty in predicting subsequent patterns arises because material and structural nonlinearities, including complex rheological laws and large-scale deformations, are  intertwined~\cite{Nakahara2006, Audoly2005, Heisser2018}.

The complexity of the crack process is vividly illustrated in the morphogenesis of fruit surfaces. 
The reticulated rind of the muskmelon or cantaloupe (\textit{Cucumis melo} L. var. \textit{reticulatus}), characterized by a complex, netted pattern of lignified tissue (Fig.~\ref{fig:1}(b)), reflects a history of crack process upon growth.
In the early growth stages, the surfaces of muskmelons are smooth, without any patterns. Approximately two weeks after pollination, the epidermis stiffens, while the inner meat expands. This differential expansion biaxially stretches the outer layer until its failure, triggering the formation of a crack network~\cite{Gerchikov2008, Akiba2022}. 

Inspired by these botanical growth processes, we investigate how surface curvature prescribes crack morphology in shell structures. In thin shells, geometric nonlinearity induces an intrinsic rigidity that fundamentally alters stress distribution~\cite{Lazarus2012, Vella2012}. While curvature is known to modulate fracture in sheets draped over rigid substrates~\cite{Mitchell2017}, predicting crack patterns on naturally curved shells remains a challenge due to the inevitable coupling of material and geometric nonlinearities.

Here, by combining tabletop experiments and phase-field-based finite element simulations, we demonstrate that geometry-oriented anisotropic stress governs the fracture morphology of curved shells. This geometric scenario is consistent with the crack pattern of young muskmelons or the geometry of the Europa Linea.

\section{Results}

\subsection{Geometry controls the crack pattern of shells}

\begin{figure}[h]
    \centering
    \includegraphics[width =\textwidth]{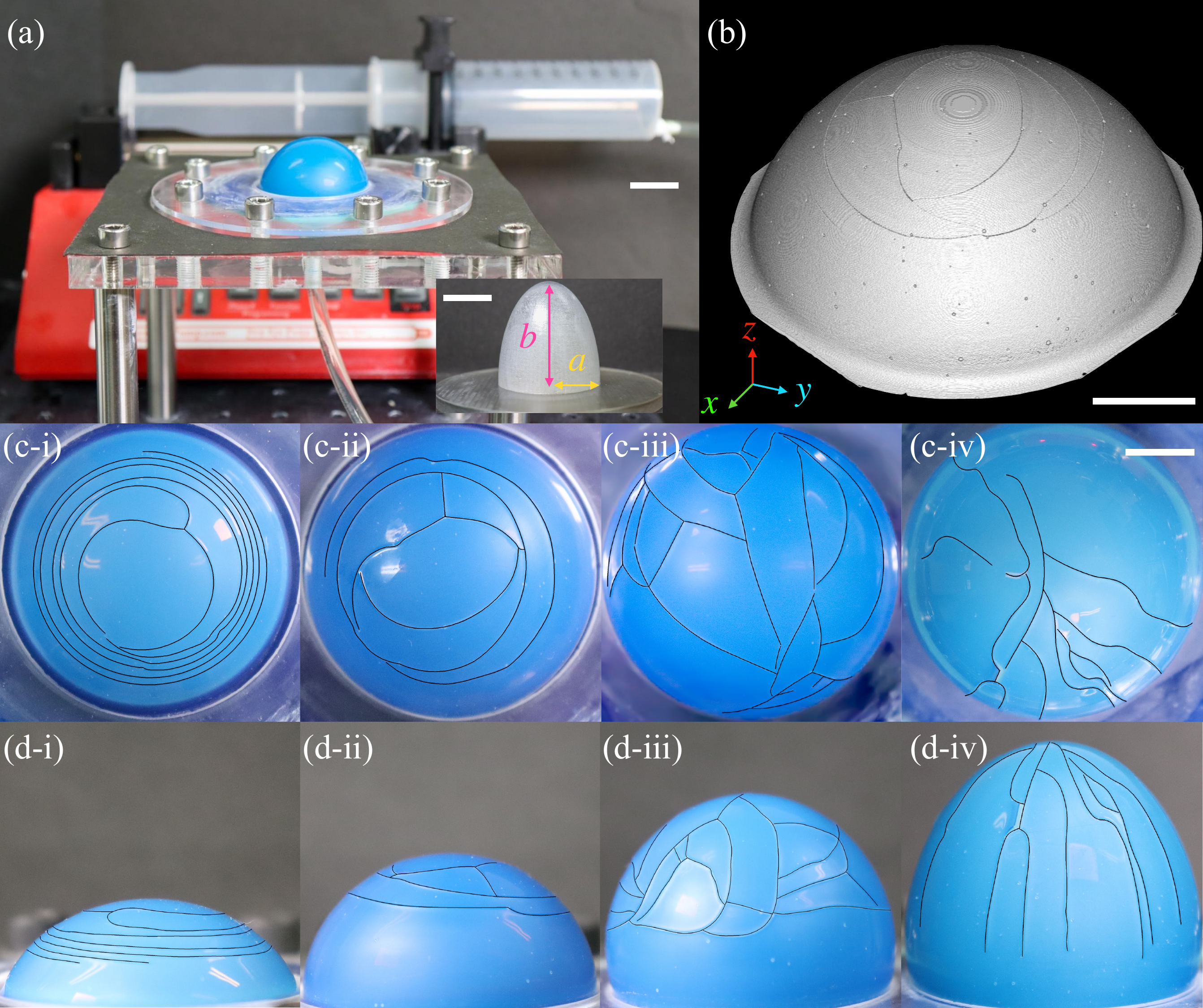}
    \caption{\textbf{Crack pattern of bilayer spheroidal shells} (a) Experimental setup of our model experiments, where our shell is pressurized from inside by a syringe pump. Inset: photograph of the inner mold of a spheroid ($b/a = 1.5$). The lateral curvature, $a$, and thickness, $h$, are fixed throughout as $(a,h) = (20, 0.4)$~mm. 
    (b) Reconstructed volumetric image through \muCT scan. The black curves represent the crack paths on the shell. 
    Crack morphology of bilayer shells: (c) top-view and (d) side-view. The photographs of shells with curvature ratios of (c-i)(d-i) $b/a = 0.5$, (c-ii)(d-ii) $b/a = 1.0$, (c-iii)(d-iii) $b/a = 1.5$, and (c-iv)(d-iv) $b/a = 2.0$. Scale bars: (a)~20mm, (b)-(d)~10mm.}
    \label{fig:2}
\end{figure}

Our model experimental system is a bilayered spheroidal shell pressurized from the inside (Fig.~\ref{fig:2}(a)). Thin bilayer shells of inner elastic and outer brittle silicone elastomer are fabricated with the coating method~\citep{Lee2016}, in which the mixed and degassed liquid elastomer is poured onto the 3D-printed spheroidal mold with two radii, $a$ and $b$ (Fig.~\ref{fig:2}(a) inset), and cures while coating the mold surface. By applying the coating method twice, we obtain thin bilayer shells with nearly uniform thickness $h\simeq0.4$~mm (0.2~mm per layer). The shell is pressurized by a syringe pump at a constant speed and is inflated. The outer surface is stretched and cracked. One of the X-ray computed tomography~($\mu$CT) images of the crack pattern is shown in Fig.~\ref{fig:2}(b), where the cracks in the $\mu$CT images appear as black curves and will be analyzed later to quantify their geometry.

The crack patterns are qualitatively different for shells with different curvature (aspect) ratios $b/a$ (Fig.~\ref{fig:2}(c)(d)). When the shell is nearly flat $b/a=0.5$, the crack pattern is almost lateral. As we increase $b/a$, the crack pattern becomes random at $b/a\approx 1$, and the crack propagates along the meridian for $b/a = 2.0$. The remarkable difference in crack patterns arises from the mechanics of the base spheroidal shells, suggesting that crack propagation could be tunable by the curvature of the substrate. 

\begin{figure}[!h]
    \centering
    \includegraphics[width =\textwidth]{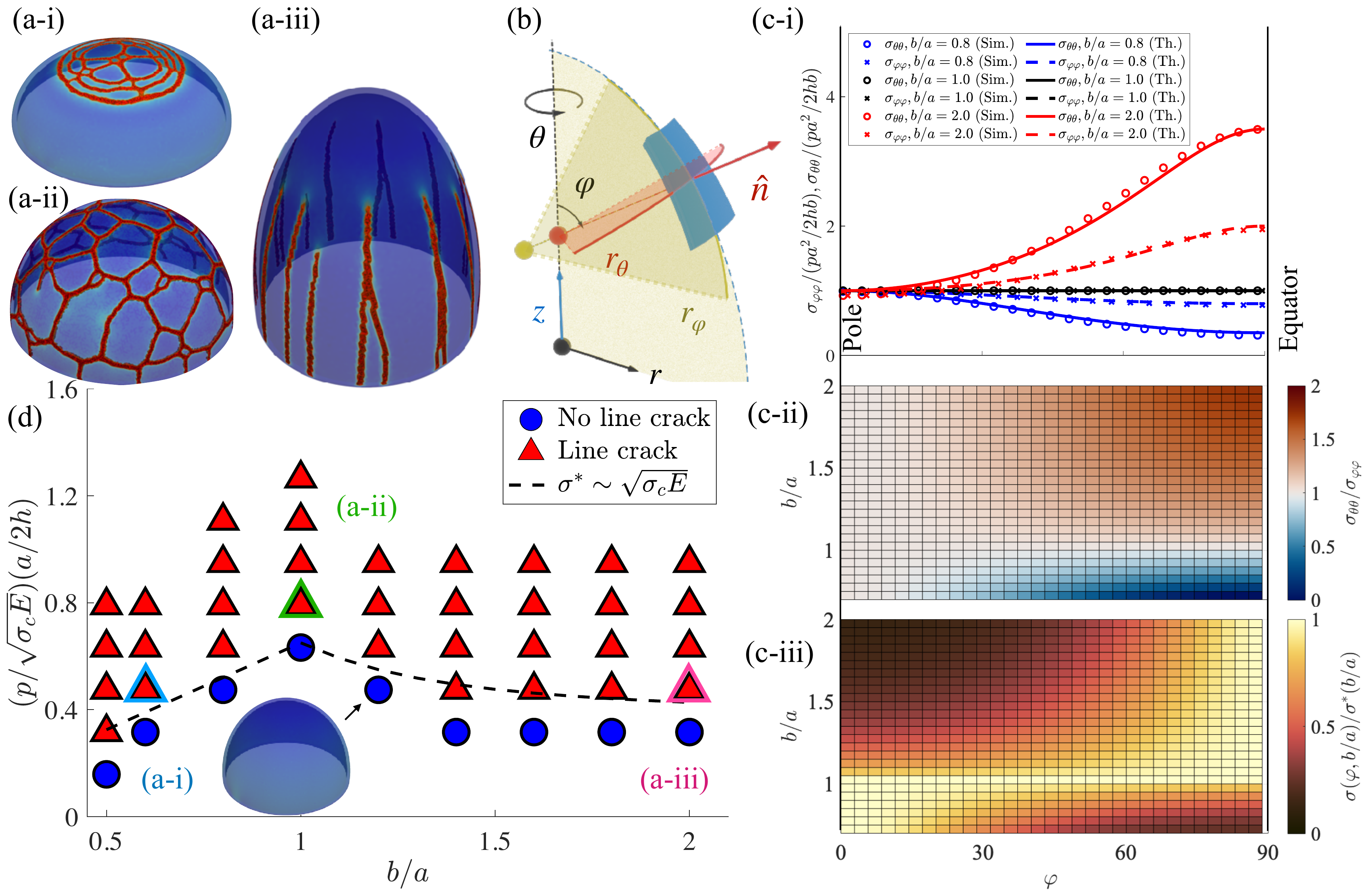}
    \caption{\textbf{Shape morphology and stress profile obtained from the simulation results.} (a) Typical FEM snapshots for (a-i) $b/a = 0.6$, (a-ii) $b/a = 1.0$, and (a-iii) $b/a = 2.0$ (See (d) for the corresponding values of internal pressure). Red colored regions represent the cracked area. 
    (b) Schematics of spheroid geometry. The meridian and lateral radii of curvatures, $(r_{\varphi},r_{\theta})$, defined through the surface normal vector, $\hat{\bm{n}}$, characterize the local geometry. The latitude of a spheroid is defined as an angle between $\hat{\bm{n}}$ and the $z$-axis.
    (c-i) Profiles of the longitudinal, $\sigma_{\phph}$ ($\circ$), and lateral components, $\sigma_{\thth}$ ($\times$), of the stress tensors obtained from FEM in the absence of cracks are compared with the analytical formulas (solid lines). Red, black, and blue data points and curves represent simulation results and theory for $b/a = 0.8, 1.0$, and $2.0$. The origin of latitude $\varphi=0$ corresponds to the north pole. (c-ii)~The ratio of lateral and longitudinal stress, $\sigma_{\thth}/\sigma_{\phph}$, is summarized as a heatmap on the $\varphi$-$(b/a)$ plane. Red and blue denote that the lateral and longitudinal components are more dominant, respectively. (c-iii)~The analytical profile of the von Mises stress, $\sigma(\varphi, b/a)$, normalized by the maximum, $\sigma^*(b/a)$, is compared for several $b/a$ and is summarized as a heatmap. The bright color represents the latitude of the largest von Mises stress for each $b/a$. (d) Phase diagram classifying whether the line cracks are visible ($\triangle$) or not ($\circ$). The dashed line represents the prediction from Eq.~\eqref{eq:smax} with a single fitting parameter, $\gamma = 0.65$, as $\sigma^* = \gamma \sqrt{\sigma_cE}$.}
    \label{fig:3}
\end{figure}

Cracks initiate at material points that are fragile, such as structural defects or stress concentrations. Our experimental shells would inevitably have defects during manufacturing (\textit{e.g.,} air bubbles). To clarify the role of geometry in crack propagation and to minimize the effects of experimental uncertainty, we perform phase-field-combined finite-element methods (FEM)~\cite{Bourdin2000, Bourdin2014, Sicsic2014, Hirshikesh2019, Wang2022} in which crack interfaces are modeled as a scalar continuous function of material points, rather than tracking the dynamics of free boundaries. The equilibrium rest shape of pressurized FEM-bilayer shells for several aspect ratios $b/a$ exhibit consistent crack morphology. When the shell is nearly flat ($b/a = 0.6$), many cracks propagate laterally, whereas they propagate along the medians for $b/a = 2$. The cracks on the spherical shell ($b/a = 1.0$) form polygons, similar to the mesh pattern on muskmelon, indicating that a bilayer shell would provide an essential physical mechanism for the reticulated rind of muskmelons (Fig.~\ref{fig:3}(a-i)-(a-iii)). 
In addition to the crack propagation angles, the FEM snapshots indicate that crack locations exhibit different trends for $b/a \lessgtr 1$. The lateral cracks appear near the north pole, while the longitudinal ones propagate near the equator.

\subsection{Mechanism behind the crack morphology}

The von Mises criterion is widely used to predict the yielding condition, from which we expect the crack initiation in our model system. 
At fracture onset, the elastic energy density (per unit volume) $\sigma^{*2}/E$, balances with the fracture energy (stress), $\sigma_c$, as $\sigma^{*2}/E\sim\sigma_c$ with Young's modulus, $E$. The von Mises stress at the crack initiation, $\sigma^*$, would depend on the geometry of the shell, $\sigma^*=\sigma^*(b/a)$, as below.

Let $(\varphi,\theta)$ denote the latitude and longitude of the shell mid-surface $(0^{\circ}\leq\varphi\leq90^{\circ}$, $0^{\circ}\leq\theta\leq360^{\circ}$). We expect that the magnitude of the corresponding principal components of (in-plane) stress tensors, $(\sigma_{\varphi\varphi}, \sigma_{\theta\theta})$, provides the precursors of crack directions and locations, assuming that the brittle layer is within elastic regime (Fig.~\ref{fig:3}(b)).  
The balance of an infinitesimal thin volume element of the axisymmetric shell yields two balance equations~\cite{ThinPlatesShell, jawad2017stress}. The circumferential balance equation is called the membrane equation, $\sigma_{\varphi\varphi}/r_{\varphi} + \sigma_{\theta\theta}/r_{\theta} = p/h$, whereas the equilibrium condition along the meridian gives $h\sigma_{\varphi\varphi} = pr_{\theta}/2$. Here, we denote the radii of the meridian and transverse curvatures, $r_{\varphi} = a^2b^2/(a^2\sin^2\varphi + b^2\cos^2\varphi)^{3/2}$ and $r_{\theta} = a^2/(a^2\sin^2\varphi + b^2\cos^2\varphi)^{1/2}$ for spheroids, respectively. 
The essential geometric quantity for the stress profile is the local ratio of the principal radii of curvature;
\begin{eqnarray}
    \tilde{\rho} \equiv \frac{r_{\theta}}{r_{\varphi}},
\end{eqnarray}
from which we express each component of the stress tensor as 
\begin{eqnarray}
    \sigma_{\phph} = \frac{pa^2}{2hb}\tilde{\rho}^{-1/2},~~\sigma_{\thth} = (2-\tilde{\rho})\sigma_{\phph}\label{eq:stprof}.
\end{eqnarray}
The analytical formula for the axisymmetric shell, Eq.~\eqref{eq:stprof}, is in excellent agreement with our FEM simulation in the absence of any fracture, as shown in Fig.~\ref{fig:3}(c-i).

The analytical formula for the stress tensors $(\sigma_{\thth}, \sigma_{\phph})$ derived and validated above captures the essential feature of the simulation results. 
First, the ratio of the stresses, ${\sigma_{\theta\theta}}/{\sigma_{\varphi\varphi}} = 2- \tilde{\rho}$, provides the direction of dominant stress, which allows us to estimate crack direction. The lateral to longitudinal stress ratio, ${\sigma_{\theta\theta}}/{\sigma_{\varphi\varphi}}$, is larger or smaller than 1 for $b/a > 1$ and $b/a < 1$, respectively as Fig.~\ref{fig:3}(c-ii). When the shell is vertically slender as $b/a>1$, the lateral stress, $\sigma_{\thth}$, dominates the longitudinal one, $\sigma_{\phph}$, as $\sigma_{\thth}/\sigma_{\phph}>1$, where the shell is preferable to fracture along the longitudinal direction. As we decrease the aspect ratio $b/a$, the lateral stress becomes less dominant, whereas the longitudinal stress becomes more significant. Hence, the crack on the nearly flat shell propagates horizontally. 
Second, the von Mises stress, $\sigma$, at latitude $\varphi$ defined as $\sigma^2 \equiv {\sigma_{\phph}^2 + \sigma_{\thth}^2 - \sigma_{\phph}\sigma_{\thth}}$, supplies the critical conditions of fracture, relevant in estimating the latitude prone to crack, is readily derived as $ \sigma = ({pa^2}/{2hb})\sqrt{({\tilde{\rho}^2-3\tilde{\rho}+3})/\tilde{\rho}}.$ The von Mises stress, $\sigma$, behaves differently for $b/a\lessgtr 1$, which monotonically increases or decreases for $b/a > 1$ and $b/a < 1$, respectively. In other words, $\sigma$ is maximum near the equator, $\varphi=90^{\circ}$, and the north pole, $\varphi=0^{\circ}$, for $b/a > 1$ and $b/a < 1$, respectively (Fig.~\ref{fig:3}(c-iii)). This is consistent with the fact that FEM shells exhibit more cracks near the north pole or equator for $b/a\lessgtr 1$ as Fig.~\ref{fig:3}(a). Lastly, the formula allows us to estimate the geometric dependence of the fracture condition. The balance of elastic and fracture stresses provides the critical condition of the fracture as $\sigma^{*2}/E\sim\sigma_c$, where the critical von Mises stress at fracture onset, $\sigma^*$, would correspond to the maximum of $\sigma$ as 
\begin{eqnarray}
    \sigma^* = \frac{pa}{2h}\left\{
\begin{aligned}
&\frac{a}{b} ~\hspace{2.9cm}(b/a\leq 1),\\
&\sqrt{\left(\frac{a}{b}\right)^2-3\left(\frac{a}{b}\right)+3}~(b/a> 1).
\end{aligned}
\right.\label{eq:smax}
\end{eqnarray}
We plot the phase boundary predicted based on Eq.~\eqref{eq:smax} as a dashed line in Fig.~\ref{fig:3}(d). The required pressure to crack the shell surface increases as $b/a$ increases up to $b/a = 1$ and decreases for $b/a>1$, indicating that the symmetric spherical geometry is stiffer against fracture than flat or elongated shells. Our prediction based on Eq.~\eqref{eq:smax} is consistent with our simulation results, in which the equilibrium shapes are classified as either with or without line cracks. 

\section{Discussion}

\begin{figure}[h]
    \centering
    \includegraphics[width =\textwidth]{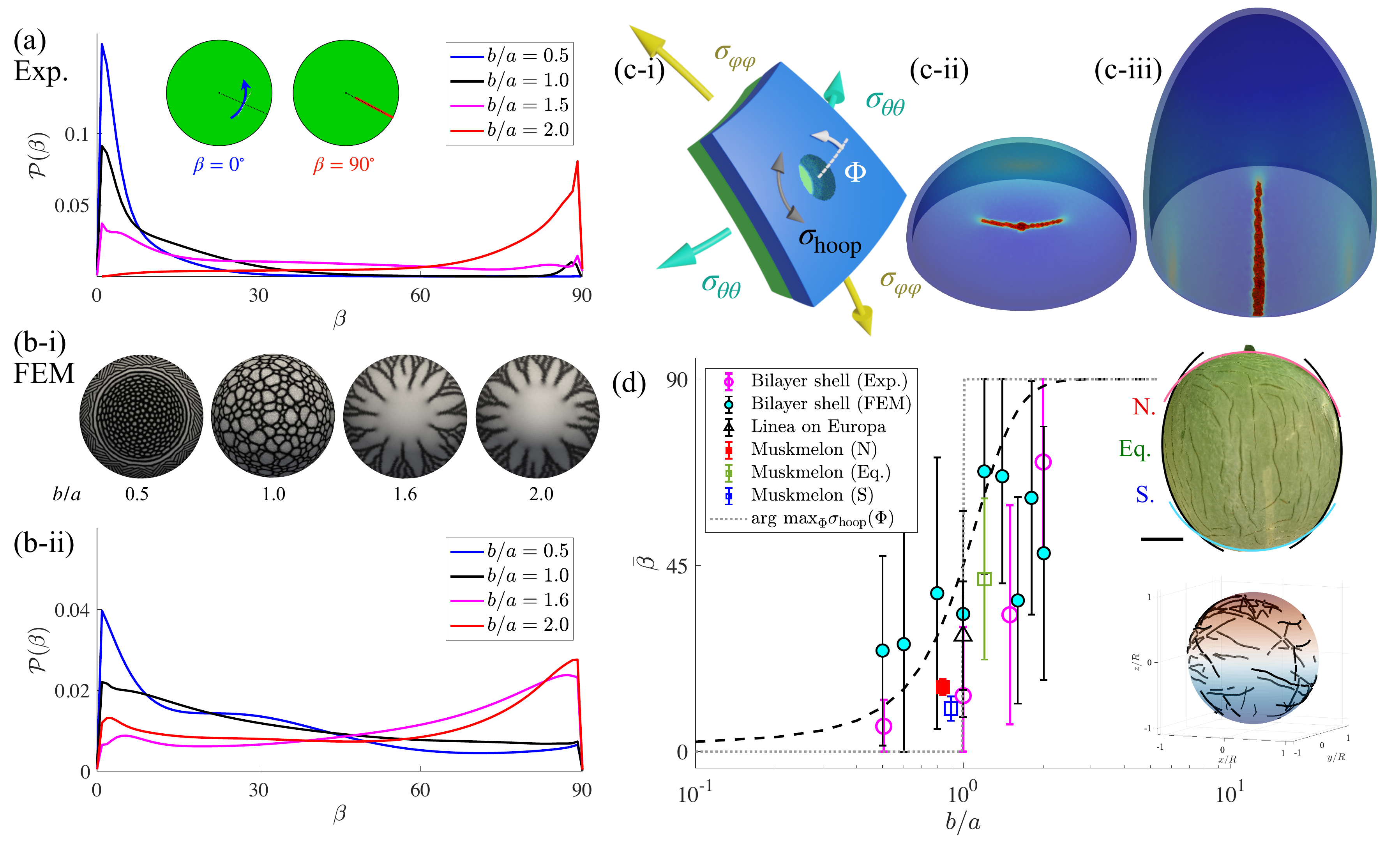}
    \caption{\textbf{Crack direction for model bi-layer shells, \linea of Europa, and young muskmelon} (a)~Crack angle distribution, $\mathcal{P}(\beta)$, for several $b/a$ of experiments. (b-i) Snapshots of FEM simulation results for developed crack patterns for each shell geometry $b/a = 0.5, 1.0, 1.6, 2.0$. The corresponding crack angle distribution is plotted in (b-ii). 
    (c-i) Schematics of the idealized crack initiation, where a single circular crack is invoked. Snapshots of FEM simulation results with an initial single defect in the vicinity of the critical pressure for $b/a = 0.8$ and 2.0 are shown in (c-ii) and (c-iii), respectively. 
    (d){Averaged crack direction $\bar{\beta}$ as a function of the curvature ratio $b/a$}. The empty and filled circular data represent the experimental and numerical results, respectively. The empty triangle is an average crack angle of the \linea of Europa, and the empty squares indicate the average crack angles in three regions of the muskmelon surface. The error bars represent the standard deviations. The insets are a photograph of a young muskmelon (scale bar: ~5 cm) and the 3D reconstruction of the \linea of Europa (normalized by its radius).}
    \label{fig:4}
\end{figure}

Our model-system study, combined with physical experiments and simulations that account for shell elasticity, indicates that nonlinear geometry governs fracture mechanics on the shell surface. Here, we focus on the crack angle, $\beta$, defined as the angle between the crack tangent and the parallel of latitude projected onto the equatorial plane (See inset of Fig.~\ref{fig:4}(a-i)), to compare crack geometry across physical systems at different length scales.

We measure crack-angle distributions of $\beta$, $\mathcal{P}(\beta)$, for our model systems using $\mu$CT scans and FEM. We find that the distribution function, $\mathcal{P}(\beta)$ in experiments, have a sharp peak around $\beta\simeq0^{\circ}$ for $b/a = 0.5$ and the peak gradually shifts to $\beta\simeq90^{\circ}$ with the increase of $b/a$~(Fig.~\ref{fig:4}(a-i)), consistent with our qualitative observations, where the lateral crack changes to random and then longitudinal as $b/a$ (Figs.~\ref{fig:2}(c)(d)). We compare the image analysis with the top view of the fully developed crack patterns obtained in FEM (Fig.~\ref{fig:4}(b-i)) and qualitatively observe the same trends shown in Fig.~\ref{fig:4}(b-ii). Despite the complexity of shell fracture processes, the statistics of cracks from experiments and FEM are in excellent agreement. 

We have shown that the crack pattern classification arises in a reproducible manner in a bilayer shell. Here, we aim to formulate the crack angle by applying fracture mechanics. Consider a tiny circular crack on the brittle layer that will initiate the crack propagation as shown in Fig.~\ref{fig:4}(c-i). The stress concentration that appears around the hole is known as the Kirsch solution~\cite{Kirsch1898}. Assuming the in-plane stressed conditions, the circumferential hoop stress at the crack surface, $\sigma_{\rm hoop}$, is computed by superposing the Kirsh solution as
\begin{eqnarray}
    \sigma_{\rm hoop}(\Phi) = (\sigma_{\phph} + \sigma_{\thth})  + 2(\sigma_{\phph} - \sigma_{\thth})\cos2\Phi\label{eq:shoop}.
\end{eqnarray}
The circumferential angle, $\Phi$, giving the maximum of $\sigma_{\rm hoop}$, would determine the initial propagation direction of the crack. Indeed, Equation~\eqref{eq:shoop} would provide a reasonable estimate for the crack angles discussed so far. When the shell is vertically slender as $b/a>1$, the lateral stress is larger than the longitudinal one as $\sigma_{\phph}< \sigma_{\thth}$. Hence, the crack will propagate along the latitude as $\Phi = 90^{\circ}$, while $\sigma_{\rm hoop}$ is maximum at $\Phi = 0$ in the case of flatter shell as $b/a < 1$ because the lateral stress is less-dominated as $\sigma_{\phph}> \sigma_{\thth}$. We show the characteristic snapshots of the FEM with an initial single defect for $b/a = 0.8$ and $b/a = 2.0$, consistent with our predictions (Fig.~\ref{fig:4}(c-ii, iii)). In summary, when a crack is initiated elsewhere in the shell, the crack angle will be either lateral or longitudinal, depending on the curvature ratio $b/a\lessgtr1$.

To further examine the idealized scenario based on $\sigma_{\rm hoop}$, we introduce the average crack angle $\bar{\beta}$ as an average over each sample. The average crack angles, $\bar{\beta}$, for experiments and simulations are plotted as a function of $b/a$, which increases with $b/a$ in Fig.~\ref{fig:4}(d).  
We compare the prediction based on the hoop stress, $\sigma_{\rm hoop}$, as a dashed line in Fig.~\ref{fig:4}(d), finding that $\bar{\beta}$ of our model system does not grow as a step-function, rather, follows the logistic-like curve (See dashed line). We expect that our prediction based on $\sigma_{\rm hoop}$ is oversimplified for fully-developed crack patterns, such as those in Fig.~\ref{fig:4}(b-i). Hence, it may be necessary to formulate the stress profile in post-cracked configurations, which we leave to future work. 
We note that many cracks appear on the shell surface, inducing a heavy-tailed distribution for $\mathcal{P}(\beta)$, which would be a reason for large standard deviations for each data point. However, most experimental and numerical results indicate that the average crack angle is larger or smaller than $\sim45^{\circ}$ when $ b/a> 1$ or $ b/a < 1$, indicating that the crack direction on the shell is still governed by the shell geometry.

The fracture orientations on developing muskmelons exhibit similar geometric dependencies. As shown in Fig.~\ref{fig:4}(d), we observe that the shape of the fruit does not fit a spheroid; rather, it comprises three distinct regions near the north and south poles and the equator. The north (N.) and south (S.) poles are vast, while the surface near the equator (Eq.) is narrow, with curvature ratios of $b/a = 0.84, 0.89$, and $1.20$, respectively, whose average crack angles are plotted in Fig.~\ref{fig:4}(d). We find that crack orientation varies with local geometry: fractures near the equator predominantly align with the median, whereas those at the poles are less frequent, consistent with our physical model and potentially providing a basis for fruit growth.

We have investigated the morphology of fractures in pressurized bilayer shells, in which crack trajectories, ranging from lateral to longitudinal, are governed by the aspect ratio of the spheroid.  
Stress analyses of these pressurized geometries, along with experiments and simulations, reveal that surface curvature serves as a geometric blueprint that effectively controls crack propagation~\cite{Nam2012, Kim2021, Thorimbert2024}.
This framework provides a physical perspective on the large-scale patterns observed in planetary systems; for instance, topographic data from Europa indicate that the average angle of the \textit{lineae} aligns with trends established by our tabletop models (Fig.~\ref{fig:4}(d)).
Although the tectonics of Europa is way complex than our model system (\textit{e.g.,} plate tectonics, tidal forces are present)~\cite{Europa}, the agreement between model and the observation of Europa implies that tabletop experiments and the mechanics of thin soft materials can even offer a relevant basis for understanding the structural evolution of geophysical architectures across disparate scales.~\cite{Katsuragi2007, Yamaguchi2011, Voigtlnder2024}.

In closing this paper, we stress the pivotal role of geometry in the morphologies of cracks on curved surfaces. We have shown that the aspect ratio, $b/a$, dictates the crack morphologies. This aspect ratio is related to the Gauss curvatures at the pole, $\kappa_{\rm pole} = b^2/a^4$, and the equator, $\kappa_{\rm eq} = b^{-2}$, as
\begin{eqnarray}
    \frac{\kappa_{\rm pole}}{\kappa_{\rm eq}} = \left(\frac{b}{a}\right)^4\label{eq:GaussRat},
\end{eqnarray}
indicating that the crack pattern on the curved surface is governed by the geometric rigidity of the shell~\cite{Lazarus2012, Vella2012}. Our findings, supplemented by the analytical model, are consistent with daily experience: we crack an egg's shell on the equator, not on the poles, and grilled sausages often break lengthwise, as well as with the fact that the curvature potential can guide crack trajectories~\cite{Mitchell2017}.


\section{Methods}\label{sec11}

\subsection{Experimental sample fabrication}

We apply the coating method on a 3D-printed spheroidal base of axis $(a,b)$ to fabricate thin bi-layer shells. Spheroidal elastic shells are made of silicone elastomer (polyvinyl siloxane, Young’s modulus $E= 0.5$ MPa, Elite Double 22, Zhermack, Italy). The horizontal radius, $a$, is fixed to be $a = 20$~mm so that the identical thickness, $h$, is realized for different vertical axes, $b$. 
The mold is coated with the release agent (Ease Release 205, Smooth-On, USA) and left for 15 minutes. 
The base and catalyst solution of Elite Double 22 is mixed in equal volume, 10~mL, using a centrifugal mixer (AR-100, Thinky Corporation, Japan) for 40 s at 2000 rpm (mixing) and another 20 s at 2200 rpm (degassing). The mixed Elite Double 22 solution is poured into the mold and left to cure for 30 minutes. The brittle elastomer (Rubber Glass, Smooth-On, USA) is coated over the cured elastic layer. The base, catalyst for Rubber Glass, and coloring agent, with weights of $7.0$, $2.3$, and $0.1$~g, are mixed for 40~s and degassed for 20~s. The Rubber-Glass solution is further mixed with the curing-accelerant (Accel-T, Smooth-On, USA) of 0.23 g for 40 s, degassed for 20 s, and is poured onto the cured elastic layer from a height of 15 cm. Note that the weight of the curing-accelerant is set to be 0.23~g to realize the thickness of both elastic and brittle layers becomes $\simeq0.2$~mm.

\subsection{Experimental protocol}

Two acrylic plates clamp the equator of the bilayer shell. The top and bottom plates are 3 mm and 10 mm thick, respectively, and sandwich the equator, secured with eight screws to seal the interior as much as possible. The interior of the shell ($b/a = 0.5, 1.0, 1.5$) is pressurized using a syringe pump (NE-1000, New Era Pump Systems Inc) with a 120 mL syringe. We use two pumps to pressurize the shell of $b/a = 2.0$. The flow rate is set to $30~ {\rm mL/min}$, and we stop pressurizing when a crack appears. This pressurizing protocol is performed for a sample of each $b/a$.  

\subsection{$\mu$CT measurement and image analysis}

We scan our experimental sample (without internal pressure), using the micro-computed tomography (XDimensus 300, Shimadzu Corporation, Japan). The reconstructed 32-bit raw data ($\simeq 20$ GB) is converted to 8-bit, and the volume is rescaled by 1/8 in ImageJ, so that our in-house code (MATLAB) can handle the volumetric data.

The reduced data is implemented into MATLAB, and the crack pattern is extracted. We binarize the image data so that the elastomer and the crack (air) appear black and white, respectively. Special care is required to obtain smooth data on the crack geometry because the crack volume is small. We apply the \texttt{imopen} function to the binary image. The post-processed image is sensitive to the choice of structuring element (SE). We use a disk-shaped SE of two different sizes to extract the voxel locations independent of SE, which are regarded as the crack locations. The binarized data would contain air bubbles, which need to be excluded for the crack analysis. To remove air bubbles from voxels, we inflate the image (\texttt{imdialate} with spherical SE of size 5). The crack voxels are connected upon inflation, while air bubbles are isolated, which could be removed by setting an appropriate threshold.

We now have voxel data for the crack paths only. To analyze the geometry quantitatively, we approximate crack paths by smooth curves. We project the extracted 3D crack trajectory onto the $x$-$y$ plane and analyze it as a set of 2D curves for ease of analysis. Note that the 3D trajectory could be reconstructed from the projected data by using the natural geometry of the spheroid. The projected curves are skeletonized to reduce the pixel data into a set of skeletons (1 pixel-width curve) with the \texttt{bwskel} function. 

The pixels in skeletonized curves must be ordered and separated into the set of curves to compute the tangent. 
We split branched cracks into curves by classifying the pixels into three (i)~edge, (ii)~branched point, and (iii)~intermediate. We count the number of white pixels surrounding each pixel. If the number of pixels is one, the pixel belongs to (i)~the edge pixels. Pixels surrounded by more than two white pixels are (ii)~branched points. The pixels belonging to (iii)~the intermediate of the curves are surrounded by two white pixels. 
Starting from one of the edge pixels, we track the trajectory of the crack (along intermediate pixels) until we reach another edge pixel or one of the branched points. We also perform the same tracking analysis starting from a branched point to identify the crack connecting two branched points. Given that the pixels belonging to each curve are ordered, we perform \texttt{smoothdata} against the pixel data, whose $x,y$ coordinates are integers, to transform them into continuous $x,y$ coordinates of the curve. The crack angles are calculated for each pixel using the smooth curves obtained as above. 

\subsection{Image analysis of muskmelon surface}
We use a muskmelon under growth (\textit{panna melon}) from the farmer in Minami-Alps city, Yamanashi prefecture, Japan, for our image analysis. The melon was hydroponically grown in the farm of Wa Fu-Ru Inc. (Yamanashi, Japan). The cross-pollination date is 4th October, 2023, and the shape is recorded on 23rd October, 2023. The data of the crack angle of the melon (Fig.~\ref{fig:4}(c)) is obtained from the photograph of the front view. The melon surface image is split into the northern and southern hemispheres. We further split each hemisphere into the region near the pole and the equator to obtain $(a,b)$. The latitude of 30\% of the arclength from the pole (in each hemisphere) is regarded as the region near the pole. 
We take several front-view photos of the melon and use them to fit an ellipse, obtaining the radii $(a,b)$. The radii of the north, south poles, and equators are respectively obtained as $(a, b) = (4.88, 4.12), (5.03, 6.04), (4.88, 4.39)$~cm (Two regions near the equators give the same $(a, b) = (5.03, 6.04)$~cm). We analyze the cracks located near the meridian of $10^{\circ}$ latitude width. We pick four meridians separated by the longitude of $90^{\circ}$ for the analysis. 

\subsection{Image analysis for the Europa surface}

Geological data for Europa are obtained from the United States Geological Survey (USGS), which provides the longitude and latitude of all named lineae ~\cite{EuropaGeo}. Using Mercator and polar stereographic maps, we reconstruct the 3D geometry of Europa's spherical surface. We assume that the shape of Europa is a sphere $b/a = 1.0$ based on the literature~\cite{Europa}.

\subsection{Phase-field combined Finite Element Simulation}

We model the fracture process of our bilayer shell based on the phase-field approach developed in \textit{e.g.,} Refs.~\cite{Bourdin2000, Bourdin2014, Sicsic2014, Hirshikesh2019, Wang2022}. The total energy of our model, $\Pi[\bm{u}, \phi] = \Pi_{\rm st}[\bm{u}, \phi] + \Pi_{\rm fr}[\phi] - W_{\rm ex}[\bm{u}]$, is given by the sum of energy functionals for the strain energy, $\Pi_{\rm st}$, fracture energy, $\Pi_{\rm fr}$, and external work, $W_{\rm ex}$, with the displacement vector, $\bm{u}$, and phase field, $\phi$. Here, the phase field, $\phi\in[0,1]$, is a scalar damage field varying between 0 (elastic material) and 1 (fully fractured). By introducing the degradation function of $\phi$ as $g(\phi) \equiv (1-\phi)^2 + k$ (a small parameter, $k=10^{-3}$, is introduced to avoid the numerical instability), the strain energy is given by $\Pi_{\rm st}[\bm{u}, \phi]  = \int_{\Omega} g(\phi)\mathcal{E}(\bm{\epsilon})dV$, where the strain energy density of an isotropic linear elastic material is denoted as $\mathcal{E}(\bm{\epsilon}) \equiv \sigma_{ij} ^{(0)}\epsilon_{ji}/2$ with the linear strain tensor $\epsilon_{ij} = (\partial_iu_j + \partial_ju_i)/2$ and the (undamaged) stress tensor, $\sigma_{ij} ^{(0)} \equiv 2 \mu \epsilon_{ij} + \lambda \mathrm{tr}(\bm{\epsilon}) \delta_{ij}$, where $\mu \equiv E/ [2(1+\nu)]$ and $\lambda \equiv E\nu /[(1+\nu)(1-2\nu)]$ are the Lam\'e constants expressed by Young's modulus $E$ and Poisson's ratio $\nu$. Note that the stress tensor under fracture, $\sigma_{ij}$, is denoted as $\sigma_{ij} = g(\phi)\sigma_{ij} ^{(0)}$. 
We adapt the fracture energy functional that realizes the Allen--Cahn-type equation for $\phi$ as $\Pi_{\rm fr}[\phi]  = (G_c/2)\int_{\Omega} \{(\phi^2/\ell_c) + \ell_c \lvert\nabla\phi\rvert^2\}dV$, where the fracture toughness, $G_c$, and characteristic length, $\ell_c$ have been introduced. 
The first term of $\Pi_{\rm fr}$ is the fracture potential, while the second term suppresses the sharp spatial gradient of $\phi$, thereby stabilizing the simulation. This fracture functional, $\Pi_{\rm fr}$, converges to the Griffith-like energy in the limit $\ell_c\to0$~\cite{Bourdin2014}. In our simulation, we set the characteristic length $\ell_c$ to be sufficiently smaller than the system size. The external work done by the pressure is written as the surface integral over the shell interior, $\mathcal{S}_{\rm in}$, as $W_{\rm ex} = - P\int_{\mathcal{S}_{\rm in}} \bm{u}\cdot\hat{\bm{n}}ds$, where $\hat{\bm{n}}$ represents the outward unit normal vector of the surface. Based on the variational calculus against $\Pi[\bm{u},\phi]$, we solve a set of two partial differential equations (PDEs); the mechanical balance equation for the displacement, $\bm{u}$ as $\partial_j\sigma_{ji} = 0$, and the Allen--Cahn equation for the phase field, $\phi$ as $\eta d\phi/dt = -g'(\phi) \mathcal{E}(\bm{\epsilon}) + (G_c/2)[\alpha'(\phi)/\ell_c + \ell_c \nabla^2 \phi]$, where $\eta$ is a coefficient with dimension of viscosity.
The latter equation is integrated implicitly in time.
The equilibrium configuration for $(\bm{u},\phi)$ is constructed by integrating these two equations for each variable alternately over a sufficiently long time interval such that the variables, $(\bm{u},\phi)$, are almost stationary in time. 

To realize the Dirichlet (displacement) boundary condition at the equator of the FEM spheroid, $z = 0$, we set the components of the displacement vector, $\bm{u}(x,y,z)$ as follows: $u_1(- a, 0, 0) = 0, u_2(\pm a, 0, 0) = 0, u_3(x,y,0) = 0$, which enforces the deformation of the shell at $z=0$ along $xy$ plane. The interior of the spheroid (internal surface of the elastic layer) is pressurized as $\sigma_{ij}n_j = - Pn_i$ on $\mathcal{S}_{\rm in}$. The phase field of the elastic layer is fixed as $\phi = 0$, while that of the brittle layer changes upon pressurization. On the outer surface of the shell (outer surface of the brittle layer), we adopt the boundary condition for $\phi$ as $(\nabla\phi)\cdot\hat{\bm{n}} = 0$. 

The procedure to integrate the evolution equations for $(\bm{u},\phi)$ is as follows. Given the pressure $P$, we update the phase field according to the Allen--Cahn equation. The updated $\phi$ is substituted into the balance equation to obtain the displacement in the next time step. We repeat this procedure until the system reaches the stationary state. Specifically, we linearly increase the pressure from 0 to $P$ within 100 discrete steps. Then, $P$ is kept constant to relax the system to equilibrium. The structure is meshed with a tetrahedral element set comprising $8\times10^4$ nodes, using the free software GMSH~\cite{Gmsh}. Although the mesh size is sufficiently small, the meshed structure can trigger a fracture, a generic feature in fracture simulations. Note that the two nodes are distributed along the thickness directions such that $\ell_c$ is comparable with the mesh size. $\bm{u}$ and $\phi$ are discretized using quadratic finite elements. The numerical analysis and integration are performed with the open source finite element software (FEniCS)~\cite{FEniCS}. We fix the following parameters: virtual time step for Allen--Cahn equation, $\Delta t = 10\eta /E$, the thickness of the bilayer shell, $h/a = 0.05$, the characteristic length, $\ell_c/a = 0.01$, Poisson's ratio, $\nu=0.49$, and the fracture toughness, $G_c/(E a) = 1$, whereas the height of the shell, $b$, and the inner pressure, $P$ are varied.

\backmatter

\bmhead{Supplementary information}
Supplementary information includes the Derivation of stress field, Pressure dependence of crack angle distribution, Crack propagation in the presence of an initial single defect, and the movies of the relaxation process in FEM.

\section*{Declarations}

\bmhead{{Acknowledgements}} This work was supported by MEXT KAKENHI 24H00299 (T.G.S.), JST FOREST Program, Grant Number JPMJFR212W (T.G.S.). 

\bmhead{Competing Interests} The authors declare no competing interests.

\bmhead{{Data Availability}} {Experimental and simulation data are available at open repository, DOI:~\texttt{10.5281/zenodo.19206747}.}

\bmhead{Code availability}
The codes for our numerical simulations are available upon request. 

\bmhead{Authors' contributions} N.S., Y.A., K.K., H.N., R.T., and T.G.S. designed the research and interpreted the results. N.S. performed experiments for bi-layer shells. Y.A. and T.G.S. performed the image analysis of muskmelons and linea on Europa, respectively. K.K. and N.H. implemented the finite element simulations, and R.T. supervised the numerical investigations. T.G.S. managed the project. Y.A., K.K., H.N., R.T., and T.G.S. wrote the paper. 



\bibliography{sn-bibliography}




\section*{Supplementary material}

# Derivation of stress field

Our theoretical basis in this paper is the stress profile of a spheroidal shell. Here, we derive the set of ordinary differential equations (ODEs) of the stress tensor, supplementing the brief derivation employed in the main text (1, 2).

We consider a shell of revolution of thickness $h$. Our shell is axisymmetric about the $z$-axis, and the distance from the $z$-axis to the shell mid-plane (along $xy$-plane) is denoted by $r$, and we describe the azimuthal angle and latitude as $\theta$ and $\varphi$ as in the main text. Note that the latitude of revolution in general is given by the angle between the surface normal vector $\hat{n}$ and the $z$-axis. We introduce the stress resultant $(N_\theta, N_\varphi)$ as the force acting on the shell cross-section per unit width as

$$N_\theta = h\sigma_{\theta\theta}, \quad N_\varphi = h\sigma_{\varphi\varphi}. \tag{S1}$$

We aim to derive the governing equation for $(N_\theta, N_\varphi)$ by considering the equilibrium condition of infinitesimal volume of the shell. Our shell of revolution is pressurized from the inside, with the inner surface subjected to pressure $p$. When the shell is thin enough, the radial component of the stress tensor decays from $\sim p$ to zero along the thickness. Hence, it is negligible compared with the in-plane stresses as $|\sigma_{rr}| \ll |\sigma_{\theta\theta}|, |\sigma_{\varphi\varphi}|$. From the axisymmetry of the shell, the off-diagonal components of the stress tensors are zero, where we consider $(N_\theta, N_\varphi)$, which depends on $\varphi$ only below (independent of $\theta$).

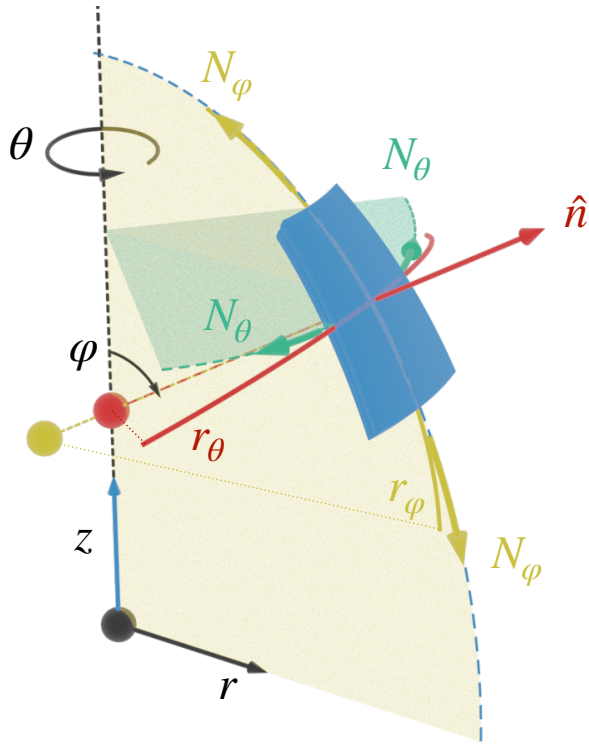

**Figure S1**: Schematics of infinitesimal volume of shell of revolution.

Consider the infinitesimal volume element near a material point at latitude and longitude $(\varphi, \theta)$ with corresponding width $(d\varphi, d\theta)$ and thickness $h$ (See blue element in Fig. S1). Let $r_\varphi$ and $r_\theta$ be

S2

the radius of meridian curvature near the material point and the distance between $O$ and the $z$ axis along the normal vector, $\hat{n}$, respectively. The former radius provides the meridional arclength of the infinitesimal element, $ds$, as $ds = r_\varphi d\varphi$. The latter radius corresponds to the radius of curvature along the lateral direction. In the case of a spherical shell, these two radii are equal: $r_\varphi = r_\theta$, while they are different for a revolution (e.g., spheroids) in general.

The force-balance equations along the median and latitude lines are derived as follows. We consider the cross-section parallel and normal to the $z$-axis of the infinitesimal element. The balance equation along the median is derived by considering the corresponding components along the median tangent. We note that the latitude component of the stress resultant, $N_\theta$, also contributes to the balance equation along the median as

$$-N_\varphi r d\theta + \left(N_\varphi + \frac{dN_\varphi}{d\varphi}d\varphi\right)\left(r + \frac{dr}{d\varphi}d\varphi\right) - N_\theta d\theta ds \cos\varphi. \tag{S2}$$

By taking the limit of $d\theta d\varphi \to 0$, we get the ODE as

$$\frac{d}{d\varphi}(rN_\varphi) - r_\varphi N_\theta \cos\varphi = 0. \tag{S3}$$

Next, the force balance equation along the surface normal vector, $\hat{n}$, by collecting all the corresponding components as

$$-N_\varphi r d\theta d\varphi - N_\theta d\theta ds \sin\varphi + pr d\theta ds = 0. \tag{S4}$$

By taking the $d\theta d\varphi \to 0$ limit, we get the second equation as

$$\frac{N_\varphi}{r_\varphi} + \frac{N_\theta}{r_\theta} = p, \tag{S5}$$

where we use the geometric relation between $r$ and $r_\theta$ as $r = r_\theta \sin\varphi$. Equation (S5) (membrane equation) allows us to erase $N_\theta$ from Eq. (S3) as

$$\frac{d}{d\varphi}\left(r_\theta \sin\varphi N_\varphi\right) + N_\varphi r_\theta \cos\varphi = p r_\varphi r_\theta \cos\varphi. \tag{S6}$$

Equation (S6) can be integrated after multiplying $\sin\varphi$ on both hand sides as

$$N_\varphi(\varphi) = \frac{p}{r_\theta \sin^2\varphi}\left(\int_0^\varphi r_\varphi r_\theta \cos\varphi \sin\varphi d\varphi + C\right), \tag{S7}$$

with the integration constant, $C$. By identifying the boundary condition and shell geometry, $(r_\varphi, r_\theta)$, we can calculate the profile of $N_\varphi$ from Eq. (S7).

**Differential geometry of spheroid**

To compute the stress field of a pressurized spheroidal shell using Eq. (S7), we need to calculate the radii of curvature of the spheroid. In this section, we summarize the differential geometry of a spheroid. The cross-section parallel to the $z$-axis of the spheroid is a part of an ellipse of radii $(a, b)$, whose equation is given by $r = (a/b)\sqrt{b^2 - z^2}$ obtained from $(r/a)^2 + (z/b)^2 = 1$ with

$r > 0$. The latitude of the spheroid relates $z$ and $r$ as $dz/dr = \tan \varphi$, from which we obtain $(r, z)$ as a function of $\varphi$ as

$$r(\varphi) = \frac{a^2 \sin \varphi}{(a^2 \sin^2 \varphi + b^2 \cos^2 \varphi)^{1/2}}, \quad z(\varphi) = \frac{b^2 \cos \varphi}{(a^2 \sin^2 \varphi + b^2 \cos^2 \varphi)^{1/2}}. \tag{S8}$$

Equation (S8) allows us to write two radii $(r_\varphi, r_\theta)$ as

$$r_\varphi(\varphi) = \frac{\{1 + (dr/dz)^2\}^{3/2}}{|d^2r/dz^2|} = \frac{a^2 b^2}{(a^2 \sin^2 \varphi + b^2 \cos^2 \varphi)^{3/2}} \tag{S9}$$

$$r_\theta(\varphi) = \frac{r(\varphi)}{\sin \varphi} = \frac{a^2}{(a^2 \sin^2 \varphi + b^2 \cos^2 \varphi)^{1/2}}, \tag{S10}$$

which will be substituted into Eq. (S7). Note that the Gauss curvature, $\kappa_G$, is obtained as $\kappa_G = 1/(r_\varphi r_\theta)$.

## Stress profile of spheroidal shells

By substituting Eq. (S9) and (S10) into Eq. (S7), we compute the profile of $(N_\theta, N_\varphi)$. First, we get $N_\varphi$ as

$$N_\varphi(\varphi) = \frac{p}{r_\theta \sin^2 \varphi} \left( \frac{a^2}{2} \frac{\sin^2 \varphi}{a^2 \sin^2 \varphi + b^2 \cos^2 \varphi} + C \right). \tag{S11}$$

The integration constant will be determined from the boundary condition of $N_\varphi$. The vertical force acting on the equator, $N_\varphi(\pi/2)$, is equal to the $z$-component of the integrated force from internal pressure, which is readily obtained as $N_\varphi(\pi/2) = pa/2$, thereby the integration constant being set to zero as $C = 0$. In summary, we get

$$N_\varphi(\varphi) = \frac{pa^2}{2} \frac{1}{(a^2 \sin^2 \varphi + b^2 \cos^2 \varphi)^{1/2}} = \frac{pr_\theta}{2}. \tag{S12}$$

The membrane equation, (S5), relates $N_\theta$ and $N_\varphi$ as

$$N_\theta = pr_\theta - \frac{r_\theta}{r_\varphi} N_\varphi = N_\varphi (2 - \tilde{\rho}), \tag{S13}$$

with the radial parameter, $\tilde{\rho}$, introduced in the text, $\tilde{\rho} = (a/b)^2 \sin^2 \varphi + \cos^2 \varphi$. Combining Eqs. (S1), (S12), and (S13), we reproduce Eq. (1) in the main text.

## Pressure dependence of crack angle distribution

In the main text, we show that the peak of the crack-angle distribution function, $\mathcal{P}(\beta)$, shifts from $0°$ to $90°$ as $b/a$ increases for specific pressures only. Here, we show the distribution functions for several values of pressure, finding that the qualitative features remain valid as we vary $p$. We plot the simulation results in Fig. S2(a)-(c), after normalizing the pressure, $p$, by the minimum pressure that exhibits linear cracks, $p^*$ (the lowest $p$ of the Linear crack for each $b/a$ in Fig.3(c)). The distribution functions, $\mathcal{P}(\beta)$, have a peak around $\beta \approx 0$ for $b/a = 0.5$ (Fig. S2(a)), they flatten for $b/a = 1.0$ (Fig. S2(b)), and peaks shifts to $\beta \approx 90°$ for $b/a = 2.0$ (Fig. S2(c)). In summary, the qualitative features of the distribution functions are still valid for different pressure values.

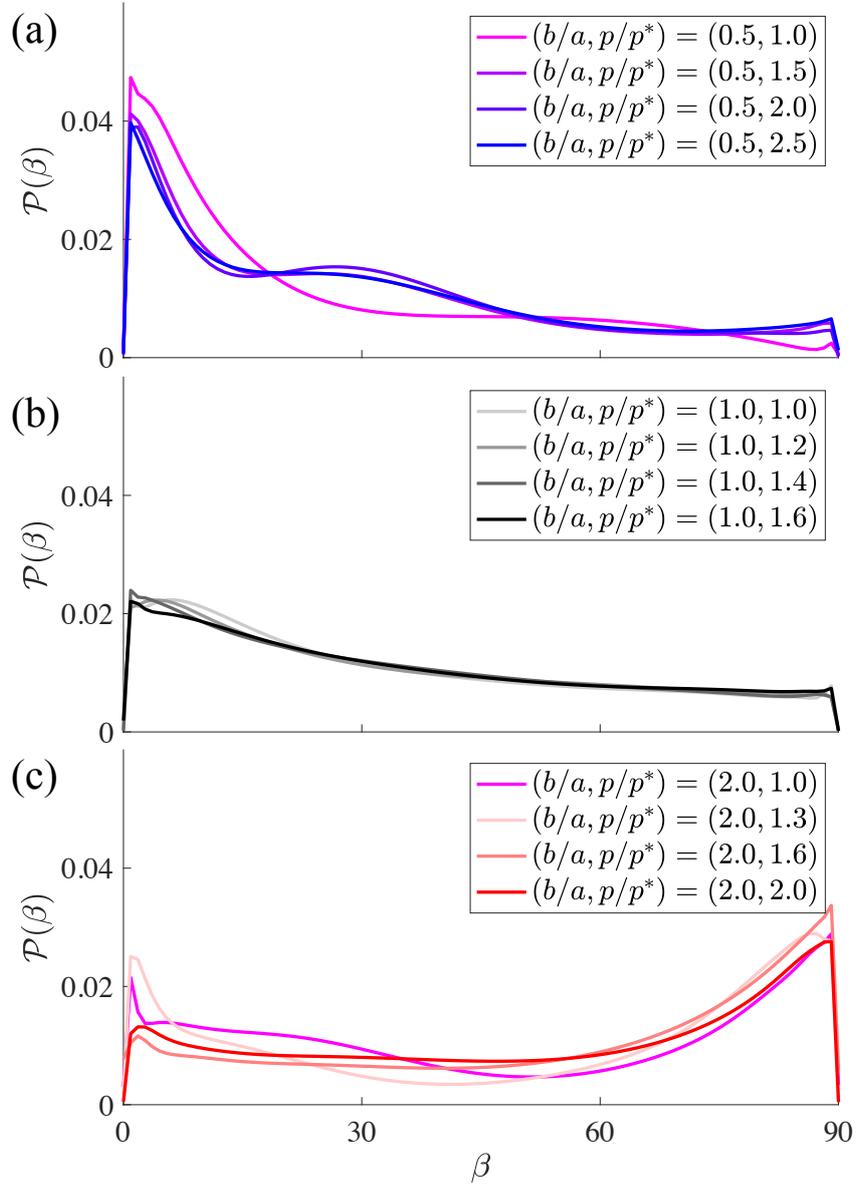

**Figure S2**: Distribution functions of crack angles, $\mathcal{P}(\beta)$, compared with different values of the pressure $p$; (a) $b/a = 0.5$, (b) $b/a = 1.0$, and (c) $b/a = 2.0$. The pressure, $p$, is normalized by the critical pressure $p^*$ for the emergence of linear crack patterns obtained in the simulation.

# Crack propagation in the presence of an initial single defect

In the main text, we show that the shell with an initial single defect exhibits either lateral or longitudinal directions, depending on the curvature ratio, $b/a$. We show that shells with defects at different latitudes exhibit similar crack morphologies. We set the initial phase field to be zero except for the initial defect. The initial defect located at the latitude $\varphi_0$ has a phase-field value of 1 (fully cracked) and is circular, with a diameter equal to the shell thickness, $h$. We integrate the time evolution under the given pressure (up to slightly above the crack initiation threshold) to reach mechanical equilibrium.

We show snapshots of $b/a = 0.8$ and 2.0 in Fig. S3. We perform the simulation with $\varphi_0 = 45°$, finding that the crack propagates horizontally and vertically for $b/a = 0.8$ and 2.0, respectively, as shown in Fig. S3(a). We note that the crack for $b/a = 0.8$ initiates at the defect, whereas the crack for $b/a = 2.0$ does not. This difference stems from the fact that crack initiation occurs at the location where the von Mises stress is highest. The stress concentration at the initial defect would be highest for $b/a = 0.8$, while the stress near the equator is highest for $b/a = 2.0$. We also show the simulation result with $\varphi_0 = 10°$ for $b/a = 0.8$ and $\varphi_0 = 80°$ for $b/a = 2.0$ in Fig. S3 (b) and (c), respectively, where the dependence of $b/a$ on crack direction is also unchanged. These comparisons imply that the initial crack directions are determined by the curvature ratio and are insensitive to the location of the defects.

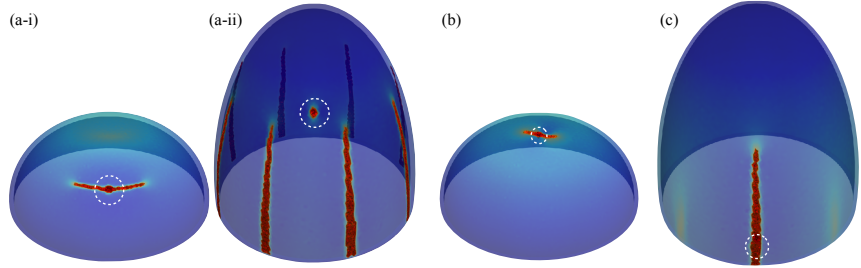

**Figure S3**: FEM snapshots of the shell with initial defects. (a-i) $(b/a, \varphi_0, p/E) = (0.8, 45°, 0.016)$, (a-ii) $(b/a, \varphi_0, p/E) = (2.0, 45°, 0.013)$ (b) $(b/a, \varphi_0, p/E) = (0.8, 10°, 0.016)$, (c) $(b/a, \varphi_0, p/E) = (0.8, 10°, 0.016)$, and (d) $(b/a, \varphi_0, p/E) = (2.0, 80°, 0.013)$ )


**Caption for Movie S1**  **Crack propagation movie of $b/a = 0.5$ in FEM.** Time evolution of this movie corresponds to the time integration of the virtual time evolution equation for $\phi$.

**Caption for Movie S2**  **Crack propagation movie of $b/a = 1.0$ in FEM** Time evolution of this movie corresponds to the time integration of the virtual time evolution equation for $\phi$.

**Caption for Movie S3**  **Crack propagation movie of $b/a = 2.0$ in FEM** Time evolution of this movie corresponds to the time integration of the virtual time evolution equation for $\phi$.



\end{document}